\documentclass[11pt,oneside]{article}
\input psfig
\begin{document}
\title{\bf Cylindrical analogue of NUT space: spacetime of a line 
gravomagnetic monopole }
\author{M. Nouri-Zonoz\thanks{Supported by a grant from the ministry of culture
and higher education of Iran.} \\
Institute of Astronomy, Madingley Road, Cambridge CB3 0HA} 
\maketitle

\begin{abstract}
Using the quasi-Maxwell form of the vacuum Einstein equations and demanding
 the presence of a cylindrically symmetric radial gravomagnetic field, we
find the solution to the Einstein equations which represents the gravity 
field of a line gravomagnetic monopole. We show that this is the
 generalization of the Levi-Civita's cylindrically symmetric static spacetime,
 in the same way that the NUT metric is the empty space generalization of the
Schwarzschild metric.
\par Some of the features of this metric as well as its relation to
 other metrics are discussed.   
\end {abstract}

\section{ Introduction}
Despite their different mathematical structures the analogy between 
gravitation and electromagnetism has proven to be fruitful. This analogy
is particularly useful in the case of gravitation when we try to relate our
3-dimensional physical 
intuition and experience on the one hand with the 4-dimensional spacetime 
events 
on the other hand. It is clear that achiveing this goal requires some sort of 
splitting of spacetime to space and time components. 
In this letter we will use this analogy  and in particular the 
quasi-maxwell form of the vacuum Einstein equations to find a solution 
which corresponds to the presence of a line gravomagnetic 
monopole.\footnote {For a discussion on NUT space and gravomagnetic monopoles
 see [8].}
In what follows we will use the $1+3$ or the threading formulation (splitting)
 of spacetimes in general relativity introduced by Landau and Lifshitz [6].
In this formulation the fundamental geometrical objects used for splitting
spacetime are timelike worldlines (threads) and so the formalism is best suited
to treating stationary gravitational fields which are characterized by the
existence of a timelike Killing vector.
\subsection { The 1+3 splitting (Threading)} Suppose that $\cal M$ is the 
4-dimensional manifold of a stationary spacetime with metric $g_{ab}$ and 
$p\in \cal M$, then one can 
show that there is a 3-dimensional manifold $\Sigma_3$ 
defined  invariantly by the smooth map [1,3] 
$$\Psi (p):{\cal M} \rightarrow \Sigma_3$$ 
 where $\Psi (p)$ denotes the orbit of the timelike 
Killing vector $\mbox{\boldmath $\xi$}_t$ passing through $p$. The 3-space 
$\Sigma_3$ along with the 3-dimensional metric $\gamma_{\alpha\beta}$ 
(defined on $\Sigma_3 $) is called the factor 
space $( {\cal M} , g)\over {G_1}$, 
 where $G_1$ is the 1-dimensional group of transformations generated by 
$\mbox{\boldmath $\xi$}_t$. One can use $\gamma_{\alpha\beta}$ to define 
differential operators on  $\Sigma_3$ in the same way that $g_{ab}$ defines 
differential operators on  $\cal M$. For example the covariant derivative of a
3-vector $\bf A$ is defined as follows
$$A^\alpha_{;\beta}=\partial_\beta A^\alpha + \lambda ^\alpha_{\gamma\beta}
A^\gamma$$
$$\hspace{3in}\alpha,\beta=1,2,3$$
$$A_{\alpha ;\beta}=\partial_\beta A_\alpha - \lambda _{\alpha\beta}^\gamma
A_\gamma$$ 
where  $\lambda ^\alpha_{\gamma\beta}$ is the 3-dimensional Christoffel
symbol cosntructed from the components of $\gamma_{\alpha\beta}$ in the 
following way
$${\lambda_{\mu\nu}^\sigma}={1\over 2}\gamma^{\sigma\eta}(\partial_ 
{\nu}\gamma_{\eta\mu}+
\partial_{\mu}\gamma_{\eta\nu}-\partial_{\eta}\gamma_{\mu\nu})$$
It has been shown that the metric of a 
stationary spacetime can be written in the following form [5]
$$ds^2=e^{-2\nu}(dx^0-A_\alpha dx^\alpha)^2-{dl}^2 \eqno (1)$$
where
$$A_{\alpha}={-g_{0\alpha}\over g_{00}}\;\;\;\;\;\;\;\;\;\;
,\;\;\;\;\;\;\;\;\; e^{-2\nu}\equiv g_{00} $$
and $${dl}^2=\gamma_{\alpha\beta}dx^{\alpha}dx^{\beta}=(-g_{\alpha\beta}+
{g_{0\alpha}g_{0\beta}\over g_{00}})dx^{\alpha}dx^{\beta}\eqno (2)$$ 
is the spatial distance written in terms of 
the 3-dimensional metric $\gamma_{\alpha\beta}$ of $\Sigma_3$.
Using this formulation for a stationary spacetime one can write the
vacuum  Einstein equations in quasi-Maxwell form as follows [8]
$${\rm div} \ {\bf B}_g = 0 \eqno (3)$$
$${\rm Curl} \ {\bf E}_g = 0 \eqno (4)$$
$${\rm div} \ {\bf E}_g = -c^{-2} \left(  \ {\textstyle {1 \over 2}}
 e^{2 \nu} B_g^2 + E_g^2 \right) \eqno (5a)$$
$${\rm Curl} \ (e^{\nu} {\bf B}_g) =   - 2c^{-3} {\bf E}_g \times 
e^{\nu} {\bf B}_g \eqno (5b)$$
 $$P^{\alpha \beta} = E_g^{\alpha; \beta} +  e^{2 \nu}
 (B^\alpha _g B^\beta _g - B^2_g \gamma ^{\alpha \beta}) +
 E^\alpha _g E^\beta _g \eqno (6)$$
where the gravitational fields are
 $${\bf E}_g = -c^2 {\bf \nabla}\nu \eqno(7)$$
 $${\bf B}_g = c^2 \ {\rm Curl} \ {\bf A}. \eqno (8)$$
and $P^{\alpha \beta}$ is the 3-dimensional Ricci Tensor constructed from
 the metric $\gamma^{\alpha \beta}$. Note that all  operations in these 
equations are defined in the 3-dimensional space with metric 
$\gamma _{\alpha \beta}$.

\subsection {Non-relativistic considerations}

Before deriving the  spacetime representing
 a line gravomagnetic monopole, a simple consideration of line magnetic
 monoploes would be useful. Suppose that a line magnetic monopole is
 stretched along the $z$-axis in cylindrical coordinates. Using  Gauss's
 law and  cylindrical symmetry the $\bf B$ field produced by this monoploe 
is given by
$${\bf B}={2Q{\bf\hat\rho}\over \rho}$$
where $Q$=const. is the monopole strength per unit length.
This field can be produced either by the potential 
${\bf A}_1={2Qz\over \rho}{\bf{\hat\phi}}$ or by 
${\bf A}_2=-{2Q\phi{\hat z}}$.
The two potentials are related through a gauge transformation
$${\bf A}_1={\bf A}_2+{\bf \nabla}\chi$$
with $\chi=2Qz\phi$. Later we will find the gravitational analogue of this
gauge transformation.
\subsection {Generalized Cylindrical solution}
Levi-Civita's static cylindrically symmetric solutions of the vacuum 
Einstein equations have the following form [7]
$$ds^2=\rho^{2m}d{t^2}-\rho^{-2m}[\rho^{2m^2}(d{\rho ^2}+d{z^2})+\rho^{2}
d{\phi^2}] \eqno (9)$$where $m$ is a constant.\footnote { For
 $m=0,1$ this is a flat metric. Also note that (9) is identical to Kasner 
solution [4].}
To find the stationary cylindrically symmetric solution for empty space 
(with a radial gravomagnetic field), using equations (1) and (9), we take the
following cylindrically symmetric form for the spatial metric
$${dl}^2=\gamma_{\alpha\beta}dx^{\alpha}dx^{\beta}=e^{2\lambda(\rho)}
(d{\rho^2}+dz^2)+{\rho^2}e^{2\nu(\rho)}d\phi^2 \eqno (10)$$
For cylindrical symmetry with a radial ${\bf B}_g$, Gauss's law gives
$${\bf B}_g ={{Le^{-2\lambda-\nu}}\over\rho}{\hat{\bf\rho}} \eqno(11)$$
where $L$=const. is the gravomagnetic monople strength per unit length.
Now substituting (7) and (11) into (5a) and (5b) we get the following
 equation for $\nu(\rho)$ from (5a)
$${\rho}^2\nu^{\prime\prime}+\rho\nu^\prime-{{e^{-4\nu}L^2}\over 2 }
=0\eqno(12a)$$ while (5b) is satisfied identically.
To construct $P^{\alpha \beta}$ from equation (6) we need the 3-dimensional
 Christoffel symbols of the metric (10) which are\footnote {The summation 
convention is being employed here.} \\ \\
$${\lambda_{\mu\sigma}^\sigma}={1\over 2\gamma}({\partial_\mu}\gamma)\; \; \;
\; \; {\lambda_{\rho\rho}^\rho}={1\over 2 }\gamma^{\rho\rho}
(\partial_\rho{\gamma_{\rho\rho}})\; \; \; \; \;
{\lambda_{zz}^\rho}={-1\over 2 }\gamma^{\rho\rho}
(\partial_\rho{\gamma_{zz}})$$
$${\lambda_{z\sigma}^\tau}{\lambda_{z\tau}^\sigma}={-1\over 2 }\gamma^{\rho\rho}\gamma^{zz}(\partial_\rho{\gamma_{zz}})^2 \; \; \; \; \; \; \; \; \;
{\lambda_{\phi\sigma}^\tau}{\lambda_{\phi\tau}^\sigma}=
{-1\over 2 }\gamma^{\rho\rho}\gamma^{\phi\phi}(\partial_\rho
{\gamma_{\phi\phi}})^2$$
$${\lambda_{\phi\phi}^\rho}={-1\over 2 }\gamma^{\rho\rho}
(\partial_\rho{\gamma_{\phi\phi}}) \; \; \; \; \; \; \; \; \; \; \;
{\lambda_{\phi\tau}^\sigma}={1\over 2 }\gamma^{\sigma\eta}
(\partial_\rho{\gamma_{\eta\tau}})$$

Using the above results and equations $(7)$ and $(11)$, we find that the
surviving components of the field equation  are
$$\lambda^{\prime\prime}-{{\lambda^\prime}\over \rho}+2{\nu^\prime}^2+
2{{\nu^\prime}\over \rho}=0 \eqno(12b)$$and
$${\rho}^2\lambda^{\prime\prime}+\rho\lambda^\prime-{e^{-4\nu}L^2\over 2} 
=0\eqno(12c)$$
where the prime denotes ${\partial \over \partial \rho}$.\\
Now we proceed to solve  equations (12) for 
  $\nu(\rho)$ and $\lambda(\rho)$.
Equation $(12a)$ can be written in the following form
$$\rho {d\over d\rho}(\rho \nu^\prime)={e^{-4\nu}L^2\over 2}$$ in terms of
the variable  $u={\rm ln}\rho$ this becomes
$${d\over du}{({{d\nu}\over du})^2}={L^2}{e^{-4\nu}}{d\nu\over du}$$
which on integration gives the following solution
$${g_{00}}\equiv {e^{-2\nu}}={2m\over L}{1\over {{\rm cosh} 
[2m {\rm ln}({\rho / c})]}}
={4m \over L}{1 \over ({\rho / c})^{2m}+
({\rho / c})^{-2m}} \eqno(13)$$
Substituting this into $(12c)$ one finds
$${e^{2\lambda}}=a{\rho^{2b}}{\rm cosh}[2m {\rm ln}({\rho / c})]
=a{\rho^{2b}}{{({\rho / c})^{2m}+({\rho / c})^{-2m}}\over 2}\eqno(14)$$
 where $m$, $a$, $b$ and $c$ are constants.
Substituting $(13)$ and $(14)$ in $(12b)$ one finds that $b=m^2$. So
 writing the metric in form (1), we have 
$$\displaylines{ds^2={2m\over L}{1\over {{\rm cosh} [2m {\rm ln}({\rho/ c})]}}
{(dt-A_\alpha dx^\alpha)}^2\hfill\cr\hfill-{L\over 2m}
{\rm cosh}[2m {\rm ln}({\rho / c})]
[{2ma\over L}{\rho^{2m^2}}(d{\rho^2}+dz^2)+{\rho^2}d\phi^2]\hfill\cr}$$
To complete the metric we need to choose a potential which produces
the field (11) through (8). One can see that the potential expression
 ${\bf A}_1\equiv A_{\phi}=Lz$ $(L=\rm const.)$ will do the job \footnote 
{In a 3-dimensional space with the metric $\gamma$, the Curl $A$ is defined 
as follows
$${(Curl A)}^\alpha ={1\over 2 \sqrt \gamma}e^{\alpha\beta\gamma}(\partial_
{\beta}A_{\gamma} - \partial_{\gamma}A_{\beta})$$ where $e_{123}=e^{123}=1$.}.
So by substituting this potential expression, the full metric becomes

$$\displaylines{ds^2={2m\over L}{1\over {{\rm cosh} (2m {\rm ln}({\rho/ c}))}}
{(dt-Lzd\phi)}^2\hfill\cr\hfill-{L\over 2m}
{\rm cosh}(2m {\rm ln}({\rho / c}))
[{2ma\over L}{\rho^{2m^2}}(d{\rho^2}+dz^2)+{\rho^2}d\phi^2]\hfill (15a)\cr}$$
\medskip
\medskip
Regaining the static cylindrically symmetric metric as $L\to 0$ requires that
$$c=c_0L^{-1\over 2m}$$ $$a={L\over 2m}$$
\begin{figure}
\psfig{figure=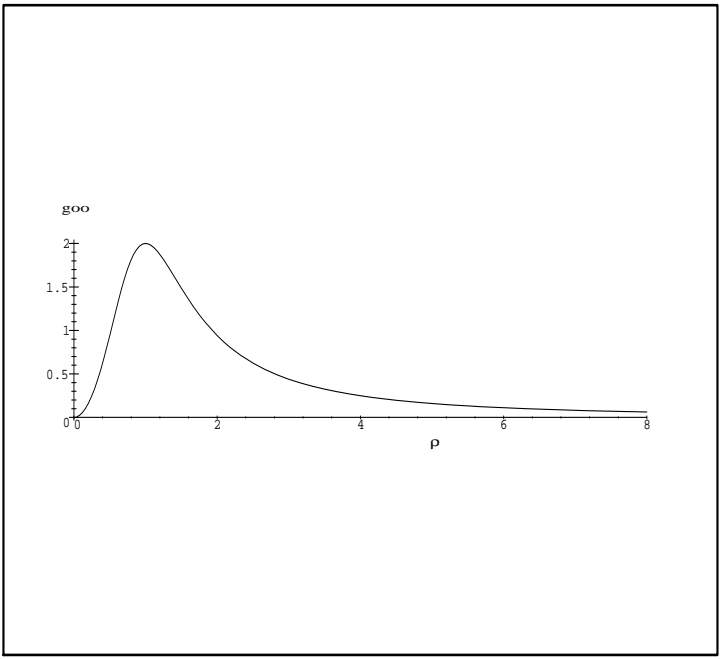,width=14cm}
\caption{$g_{00}$ as a function of $\rho$ for $L=m=c=1$.}
\end{figure}

The behaviour of $g_{00}$ as a function of $\rho$ is shown in figure 1. 
This spacetime is one of the members of the Papapetrou class [9] and has
 two event horizons, one at $\rho=0$ and another at $\rho=\infty$. 
\section {symmetries}
If we shift the $z$ coordinate in (15a) by a constant $c$ we have
$$z\to {z^\prime}=z+c$$
and $$ds^2 \to ds^{\prime 2}= e^{-2\nu(\rho)}{[dt-(Lz+b)d\phi]}^2-
e^{2\lambda(\rho)}(d{\rho^2}+dz^2)-{\rho^2}e^{2\nu(\rho)}d\phi^2$$ 
where $b=Lc$. But now 
one can shift the zero point of the time coordinate in the following 
way $$t\to {t^\prime}=t-b \phi$$
so that 
$$ds^{\prime 2}=e^{-2\nu(\rho)}{(dt^\prime -Lz d\phi)}^2-e^{2\lambda(\rho)}
(d{\rho^2}+dz^2)-{\rho^2}e^{2\nu(\rho)}d\phi^2$$
has the same form as $(15a)$ in  terms of the new 
coordinate $t^\prime$. This shows that $(15a)$ is 
translationally invariant along the $z$-direction. In other words, 
although the metric is not from the mathematical point of view cylindrically
 symmetric (i.e. not all
the components are $z$-independent), it is so physically. This  
is also clear if one represents the spacetime (15a) 
by its gravoelectric and gravomagnetic 
fields which are both explicitly cylindrically symmetric.\\ 
One can equally well choose as the gravomagnetic
potential the expression ${\bf A}_2 \equiv A_z=- L\phi$ which leads to the 
same gravomagnetic field ${\bf B}_g$. But the metric becomes
$$\displaylines{ds^2={2m\over L}{1\over {{\rm cosh} (2m {\rm ln}({\rho / c}))}}
{(dt+L\phi dz)}^2\hfill\cr\hfill-{L\over 2m}
{\rm cosh}(2m {\rm ln}({\rho / c}))
[{2ma\over L}{\rho^{2m^2}}(d{\rho^2}+dz^2)+{\rho^2}d\phi^2]\hfill (15b)\cr}$$
This metric is explicitly cylindrically symmetric but it suffers 
from multivaluedness.
If we apply the same reasoning as that we used  for the metric $(15a)$ above, 
we can associate this multivaluedness with
different choices of the zero point of the time coordinate through the 
following transformation
$$t\to {t^\prime}=t+bz\ \ \ ,\ \ \ \ \ \ \ b=2\pi nL\ \ \ \ \ n=1,2,3,...$$
As in the case of the magnetic monopole (section 1.2) the two 
different choices for the vector potential $\bf A$  correspond to
 the following gauge transformation between them
$${\bf A}_1={\bf A}_2+{\bf \nabla}\chi$$ where $\chi= Lz\phi$.
\subsection {Killing vectors}
It is already clear that $\mbox{\boldmath $\xi$}_t=(1,0,0,0)$ and
$\mbox{\boldmath $\xi$}_\phi=(0,0,0,1)$ are killing vectors of $(15a)$ however 
one can see that the form $(15b)$ can be reached alternatively by the 
transformation $t \to t^\prime = t - Lz\phi$. Now, as is obvious from 
$(15b)$, $\xi_{z}^{\prime\beta}=(0,0,1,0)$ is  a Killing vector in 
the transformed coordinates and upon converting to the previous coordinates 
we find  that 
$$\xi^\alpha ={\partial x^\alpha \over \partial x^{\prime\beta}}
\xi^{\prime\beta}=({\partial t\over \partial z^\prime} ,0,
{\partial z\over \partial z^\prime} ,0)=(L\phi,0,1,0)$$
is also a Killing vector of $(15)$.\footnote {I am grateful to 
D. Lynden-Bell and J. Katz for pointing out this to me.} 
This is a multivalued Killing vector 
whose  interpretation  might require the consideration of not just the space 
itself but also the covering space with the many values of $\phi$ unroled. 
\section { Relation to other metrics}

To find the connection between this metric and  some other solutions, we 
use the Ernst formulation of axially symmetric spacetimes. In this 
formulation the vacuum field equations, in the Weyl's coordinates 
$(\rho,z)$,\footnote {In these coordinates the metric of an axially symmetric 
stationary spacetime can be written in the following form
$$ds^2=e^{2U}(dt+Ad\phi)^2-e^{-2U}[e^{2K}(d\rho^2+dz^2)+\rho^2 d\phi^2].$$}
 can be written in the following form [2,5]
$$(\varepsilon+ \varepsilon^*)(\varepsilon_{,\rho\rho}+\rho^{-1}\varepsilon
_{,\rho}+\varepsilon_{,zz})=2({\varepsilon_{,\rho}}^2+{\varepsilon_{,z}}^2)
\eqno (16)$$where $\varepsilon$ is a complex potential defined as follows
$$\varepsilon=e^{2U}+i\omega \;\;\;\;\;\;\;\; ,\;\;\;\;\;\;\;\;\; 
 e^{2U}=g_{00}\eqno (17)$$
Every solution of $(16)$ gives a stationary axially symmetric metric. one can 
 easily see that the metric $(15a)$ is a solution of (16) with
$$e^{2U}=g_{00}={2m\over L}{1\over {{\rm cosh} (2m {\rm ln}({\rho / c}))}}$$
and $$\omega={2m\over L}{\rm tanh}(2m {\rm ln}({\rho / c}))$$
Now the following theorems lead us to some other solutions [5]
\par $(I)$. Given a stationary axisymmetric vacuum solution
 $(\varepsilon=e^{2U}
+i\omega)$, the substitution
$$U^ \prime =-U+{1\over 2} ln\rho$$ $$\omega ^\prime =i\omega$$ yields another 
vacuum solution $({U^\prime},\omega^\prime)$.
\par $(II)$. and the substitution$${U^\prime}=2U$$  $${\omega^\prime}
=i\omega$$ yields a solution
 $({U^\prime},\omega^\prime)$ of the Einstein-Maxwell equations.
\par $(III)$. For a stationary axisymmetric vacuum solution $(U,\omega),$
(in Weyl's coordinates) one obtains a corresponding cylindrically 
symmetric Einstein-Maxwell field $(U^\prime , \omega^\prime)$ by 
the substitution $$t\to iz\;\;\;\;\;\; z\to it \;\;\;\;\;\;2U\to U^\prime \;\;
\;\;\;\; \omega\to \omega^\prime$$
For example applying (I) followed by (III) to (15a), one obtains the 
Rahdakrishna metric [10], which is a time dependent cylindrically symmetric 
solution of Einstein-Maxwell equations.
\subsection {Gravitational duality rotation}

One can also obtain metric (15) by applying the gravitational duality
 rotation
(sometimes called Ehlers' transformation)\footnote {This transformation
 was first  introduced by Ehlers [1] and later developed by Geroch [3].}to 
metric (9). This transformation, in its original formulation, states that if 
$$g_{\mu\nu}=e^{2U}{(dx^0)}^2-e^{-2U}{dl}^2$$
is the metric of a static exterior spacetime, then

$$\bar g_{\mu\nu}=(a{\rm cosh}(2U))^{-1}({dx^0}-{A_\alpha}dx^{\alpha})^{2}-
a{\rm cosh}(2U){dl}^2$$
[with $a=\rm const.> 0$ , $U=U({x^\alpha})$ and $A_{\beta}=
A_{\beta}(x^{\alpha})$] would be the metric of a stationary exterior 
spacetime provided that ${A_\alpha}$ 
satisfies
$$-a {\sqrt\gamma}\epsilon_{\alpha\beta\eta}U^{,\eta}=A_{[\alpha,\beta]}
\eqno(18)$$
where $\epsilon_{\alpha\beta\gamma}$ is the flat space alternating symbol.
From equations (9) and (15a) we see that in this case
$$a={L\over 2m}$$
$${A_\alpha} \equiv {A_\phi}={L_z}$$
$$U=m{\rm ln}\rho \eqno(19)$$
and
$${dl^2}=\rho^{2m^2}(d{\rho ^2}+d{z^2})+\rho^{2}d{\phi^2}$$\\
It can easily be seen that equations (19) satisfy (18). In the same way 
the NUT metric has been shown to be  the gravitational dual of Schwarzschild
 metric [3].
\section*{ACKNOWLEDGEMENTS} 
I would like to thank my supervisor, Professor D. Lynden-Bell, without 
whose help this work would not have been done. I am also grateful to 
Dr. H. Ardavan for useful discussions and comments.

\end{document}